\documentclass[preprint,showpacs,preprintnumbers]{revtex4}

\usepackage{amsmath}
\usepackage{graphicx}% Include figure files
\begin{document}

\title{Optomechanical coupling between
a moving dielectric sphere and radiation fields:
a Lagrangian-Hamiltonian formalism}

% Force line breaks with \\

\author{H. K. Cheung and C. K. Law}
\affiliation{Department of Physics and Institute of Theoretical
Physics, The Chinese University of Hong Kong, Shatin, Hong Kong SAR,
China}

\renewcommand\Im{\operatorname{Im}}
\renewcommand\Re{\operatorname{Re}}

\date{\today}
\begin{abstract}
We present a Lagrangian-Hamiltonian formalism of a moving dielectric sphere
interacting with radiation fields. By including the interaction up to
the first order in the speed of the sphere, we derive the
Hamiltonian and perform quantization of both the field and the
mechanical motion of the sphere. In particular, we show how
independent degrees of freedom can be consistently identified
under the generalized radiation gauge via instantaneous mode
projection. Our Hamiltonian indicates the form of coupling due to velocity-dependent interactions beyond adiabatic approximation.
In addition, the Hamiltonian predicts that a geometrical quantum phase can be gained by the sphere moving in a light field.
\end{abstract}


\maketitle
%======================================================
%                     Introduction
%======================================================
\section {Introduction}

The coupling between an optically levitated sphere and cavity
fields has been a subject of research interests recently
\cite{kimble nanosphere,raizen,cirac,cirac2}. Such a system
corresponds to a basic configuration in cavity optomechanics which
explores quantum phenomena arising from the interaction between
mechanical and optical degrees of freedom \cite{review}. In
particular, since mechanical systems such as a dielectric sphere
or moving mirror have masses much greater than that of an atom,
cavity optomechanics could open a door for testing quantum theory
in macroscopic scale \cite{Bouwmeester2003,cirac2},
as well as probing Planck-scale physics~\cite{kim}. Recently,
progress in experiments has been made in demonstrating cooling and
trapping of a dielectric sphere by optical fields \cite{raizen}.

In this paper, we present a Lagrangian and Hamiltonian formalism
of the interaction of a moving dielectric sphere in quantized
radiation fields. Previously, Chang {\it et al.} \cite{kimble
nanosphere} and Romero-Isart {\it et al.} \cite{cirac} have
studied the (adiabatic) interaction Hamiltonian of the system to
zeroth order in $v$, where $v$ is the speed of the sphere. Within
this accuracy, the interaction takes the usual form $-\frac{1}{2}
\int {\bf P} \cdot {\bf E} ~d^3x$ (with the polarization ${\bf P}$
proportional to the electric field ${\bf E}$) as if the dielectric
were stationary. While such a treatment is suitable for a trapped
sphere at low temperatures, it is interesting to ask for a general
theory that takes into account interactions due to the motional
states of the sphere. In particular, it is known that a moving
dielectric possesses a velocity dependent polarization and
magnetization \cite{Landau}: ${\bf P} = (\epsilon-1)\left({\bf E}
+ {\bf v}\times{\bf B}\right)$ (to first order in $v$),
${\bf M} = -{\bf v} \times {\bf P}$, and these motion-induced
quantities could modify the dynamics of the sphere and the field.
In order to account for such effects nonrelativistically, one has
to keep the interaction part of the Lagrangian at least up to
first order in $v$, and this is our task in this paper -- to set up a
consistent Hamiltonian leading to the quantization of the full
system.

We point out that the moving sphere is a higher dimensional system
due to the translational and rotational motion in three
dimensions, and it requires a treatment different from that in
one-dimensional optomechanical systems \cite{kit}. For example, an
optical field can affect the rotational motion of a dielectric via
the electromagnetic torque exerting to it
\cite{Marton,Lee,Loudon}, and this has been studied in a sequence
of experiments \cite{Beth,Padgett,Padgett2,Dunlop,Paterson}.
Indeed, by including the rotation degrees of freedom of the sphere
in the Lagrangian, we will show that there is a kind of rotational
optomechanical coupling caused by the velocity-dependent
interactions. There is also some theoretical subtlety in
identifying independent degrees of freedom subjected to the
generalized transverse gauge. Such a gauge is commonly used for
field quantization in the presence of a stationary dielectric
medium \cite{glauber, knight, knoll}. However, for a moving
dielectric, the generalized transverse gauge somehow mixes the
mechanical degrees of freedom with the field. In this paper, we
resolve the problem by using instantaneous normal-mode projection,
and the quantization scheme is compatible with the generalized
transverse gauge.

\section {Lagrangian}\label{sec: L & EOM}

The system under investigation is depicted in Fig. 1 in which a
rigid dielectric sphere of mass $m$ and radius $R$ moves in an
electromagnetic field. The sphere is free to rotate about any axis
through its center of mass (c.m.) at ${\bf q}$. In our study, we
assume a nondispersive and nonabsorptive linear dielectric. When
the sphere is stationary, its dielectric permittivity is given by
\begin{equation}
\epsilon(\bf r,\bf q) =
 \left\{ \begin{array}{ll}
            n^2,& |{\bf r} - {\bf q}| \leq R \\
            1,& \mbox{otherwise}.
          \end{array}
 \right.
\end{equation}
where the convention $\epsilon_0 = \mu_0 = 1$ and $c=1$ is used
for convenience. In addition, the sphere is assumed nonmagnetic
in its rest frame.

\begin{figure}[h]
\centering
\includegraphics[width= 9 cm]{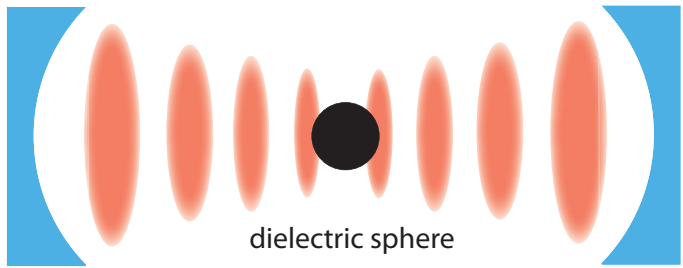}
\caption{(Color online) An illustration of an optomechanical
system involving a dielectric sphere moving freely in
electromagnetic fields. The fields can be contained in a generic
optical cavity or in free space.}
\end{figure}
The (nonrelativistic) Lagrangian of the sphere and
electromagnetic fields is given by
\begin{equation}
L = \frac{1}{2}m\dot{\bf q}^2 + \frac{1}{2} I {\boldsymbol
\omega}^2 + \int d^3 r {\cal L}(\bf r),
\end{equation}
where $I$ and ${\boldsymbol \omega}$ are the moment of inertia and
angular velocity of the sphere, respectively. The orientation of
the sphere is specified by three Euler angles $\alpha$, $\beta$, and
$\gamma$ \cite{euler ang}, and hence ${\boldsymbol \omega}$ can be
explicitly expressed as ${\boldsymbol \omega} = (\dot{\gamma}
\sin\beta \cos\alpha - \dot{\beta}\sin\alpha) \hat{\bf e}_x
+(\dot{\gamma} \sin\beta \sin\alpha + \dot{\beta}\cos\alpha)
\hat{\bf e}_y + (\dot{\alpha} + \dot{\gamma} \cos\beta) \hat{\bf
e}_z$, where $\hat{\bf e}_l$ ($l=x,y,z$) are the basis vectors. ${\cal L}(\bf r)$ is the Lagrangian density of the field after
eliminating the electronic degrees of freedom of the dielectric.
To obtain the form of ${\cal L}$, we go to an {\em inertial} frame
$S'({\bf r})$ in which the dielectric element at $\bf r$ is
instantaneously at rest. Assuming the acceleration of the
dielectric does not change its macroscopic properties, the field
Lagrangian density at $\bf r$ in $S'({\bf r})$ is given by the
familiar form,
\begin{equation}
{\cal L}' = \frac{1}{2}\left( \epsilon {\bf E}'^2 - {\bf B}'^2
\right), \end{equation} where ${\bf E}'$ and ${\bf B}'$ are the
electric and magnetic fields in $S'({\bf r})$, respectively. As
the Lagrangian density is Lorentz invariant, ${\cal L}$ can be
readily obtained from the Lorentz transformation of the fields
from $S'({\bf r})$ to the laboratory frame $S$.

In this paper we confine ourselves to a nonrelativistic motion of
the sphere, so that the velocity ${\bf v}({\bf r}) = \dot{\bf q} +
{\boldsymbol \omega}\times({\bf r} -{\bf q})$ of the dielectric
element at any point $\bf r$ satisfies $v \equiv |{\bf v}({\bf
r})|\ll c$. By keeping terms up to first order of $v$ in ${\cal
L}$, the Lagrangian reads
\begin{eqnarray}
L = \frac{1}{2}m\dot{\bf q}^2  + \frac{1}{2} I {\boldsymbol
\omega}^2 + \frac{1}{2} \int d^3 r \left(\epsilon {\bf E}^{2} -
{\bf B}^{2}\right) - \dot{\bf q} \cdot {\bf \Lambda} -
{\boldsymbol \omega}\cdot{\bf \Gamma}, \label{eq:L in terms of
omega}
\end{eqnarray}
where
\begin{eqnarray}
&& {\bf \Lambda} = \int d^3 r (\epsilon-1) \left({\bf
E}\times{\bf B}\right), \\
&&{\bf \Gamma} = \int d^3  r (\epsilon-1)({\bf r} - {\bf q})
\times\left({\bf E}\times{\bf B}\right) \end{eqnarray} are
defined. The Lagrangian (\ref{eq:L in terms of omega}) is a
generalization to that of Barton and Eberlein~\cite{Barton} and
Salamone~\cite{salamone}, which consider a one-dimensional
configuration and focus only on the uniform c.m. motion of a
dielectric slab. Here we will take both ${\bf q}$ and
${\boldsymbol \omega}$ as dynamical degrees of freedom which
interact with the field through the $- \dot{\bf q} \cdot {\bf
\Lambda} - {\boldsymbol \omega}\cdot{\bf \Gamma}$ term. These
velocity dependent interaction terms are not considered in
previous studies \cite{kimble nanosphere,cirac}. Interestingly,
${\bf \Lambda}$ and ${\bf \Gamma}$ are proportional to the linear
momentum and angular momentum of the field inside the sphere.

%\section{Generalized radiation gauge and instantaneous modes projection}

The Lagrangian should be expressed in terms of the scalar
potential $V$ and vector potential ${\bf A}$ via the substitution:
${\bf E} = -\partial_t {\bf A} - \nabla V$ and ${\bf B} =
\nabla\times {\bf A}$. In this paper we shall fix the potentials
by using the {\em generalized radiation gauge}:
\begin{equation}
\nabla\cdot\left[\epsilon({\bf r}, {\bf q}) {\bf A}\right] = 0.
\label{eq:gauge}
\end{equation}
Such a gauge condition has been employed for performing field
quantization in the presence of stationary dielectric media
\cite{glauber, knight, knoll}. It has the advantage that if the
dielectric is at rest, the scalar potential $V$ is exactly zero.
For a moving dielectric, it can be shown that $V$ under
condition (\ref{eq:gauge}) contributes an interaction term of
order $v^2$ in the Lagrangian, and hence the effects of $V$ can be
consistently neglected in our nonrelativistic Lagrangian which
keeps interaction terms up to the first order in $v$.

It is important to note that the generalized gauge
(\ref{eq:gauge}) acts as a constraining equation of the vector
potential ${\bf A}$ for a given c.m. position of the sphere ${\bf
q}$. Therefore ${\bf A}$ and ${\bf q}$ cannot be treated as
independent degrees of freedom under Eq. (\ref{eq:gauge}). To
overcome the difficulty, we identify the independent degrees of
freedom via the mode expansion of ${\bf A}$:
\begin{equation}
{\bf A}({\bf r},t) = \sum_k Q_k(t) {\bf u}_{k}[{\bf r},{\bf q}(t)],
\end{equation}
where  $\{{\bf u}_{k}({\bf r},{\bf q})\}$ is a set of normal-mode
functions of the field when the sphere is sitting at rest at the
c.m. position $\bf q$. Specifically, the mode function ${\bf u}_{k}$
with the mode frequency $\omega_k^2({\bf q})$ is defined by
\begin{eqnarray}
\nabla\times\left(\nabla\times {\bf u}_k\right) - \epsilon({\bf
r},{\bf q})\omega_k^2({\bf q}) {\bf u}_k = 0,
\end{eqnarray}
subjected to suitable boundary conditions and
$\nabla\cdot\left[\epsilon({\bf r},{\bf q}){\bf u}_k({\bf r},{\bf
q})\right] =0$ from the gauge condition (\ref{eq:gauge}). In
addition, these modes are orthonormal according to,
\begin{equation}
\int d^3 r ~\epsilon({\bf r},{\bf q}) {\bf u}_k({\bf r},{\bf q})
\cdot {\bf u}_j({\bf r},{\bf q}) = \delta_{kj}.
\end{equation}
Note that we have considered ${\bf u}_{k}$ as real functions for
convenience. For a moving sphere, ${\bf q}$ is a function of time,
therefore ${\bf u}_{k}$ can be interpreted as an {\em instantaneous normal-mode} of the field.

By projecting ${\bf A}$ onto instantaneous normal-modes, the gauge
condition (\ref{eq:gauge}) is automatically satisfied and we can
treat $Q_k$ and ${\bf q}$ as generalized coordinates, i.e., independent
degrees of freedom. The Lagrangian (\ref{eq:L in terms of omega})
becomes
\begin{eqnarray}
L &=& \frac{1}{2}m\dot{\bf q}^2 + \frac{1}{2} I {\boldsymbol
\omega}^2
+ \frac{1}{2} \sum_k \left[\dot{Q}_k^2 - \omega^2_k({\bf q})Q_k^2\right] \nonumber \\
&& - \dot{\bf q}\cdot\sum_{kj} {\boldsymbol \eta}_{kj} \dot{Q}_k
Q_j - {\boldsymbol \omega}\cdot\sum_{kj} {\bf g}_{kj} \dot{Q}_k
Q_j, \label{eq:mode L}
\end{eqnarray}
with corrections on the order $O(v^2)$. Here the
${\boldsymbol \eta}_{kj}$ and ${\bf g}_{kj}$ are coefficients
depending on ${\bf q}$:
\begin{eqnarray}
{\boldsymbol \eta}_{kj}({\bf q}) &=& -\int d^3 r [ \epsilon
\sum_{l=x,y,z} \left({\bf u}_k \cdot \hat{\bf e}_l\right)
\nabla_{\bf q} \left({\bf u}_j \cdot \hat{\bf e}_l\right) +
\left(\epsilon-1\right) {\bf u}_k \times\left(\nabla\times{\bf
u}_j\right) ],
\label{eq: eta kj}
\\
{\bf g}_{kj}({\bf q}) &=& -\int d^3 r \left(\epsilon-1\right)
\left({\bf r}-{\bf q}\right)\times\left[{\bf u}_k
\times\left(\nabla\times{\bf u}_j\right) \right]
\label{eq: g kj}.
\end{eqnarray}

We remark that validity of $L$ in Eqs.~(\ref{eq:L in terms of
omega}) and (\ref{eq:mode L}) requires that interaction terms
involving $v^2$ and higher orders are negligible. Those $v^2$
terms omitted in the Lagrangian are corrections to translation and
rotational kinetic energies such as $\delta m_{ij} \dot q_i \dot
q_j$ and $\delta I_{ij} \omega_i \omega_j$, where $\delta m_{ij}$
and $\delta I_{ij}$ are respectively  the mass tensor and
rotational inertial tensor due to the field energy stored inside
the sphere. Because $\delta m_{ij} $ is generally time dependent,
it contributes a velocity dependent force $ \delta \dot m_{ij}
\dot q_j $. Similarly, there is an angular velocity dependent
torque $\delta \dot I_{ij} \omega_j$ coming from the rate of
change of $\delta I_{ij}$. If the sphere's motion is consistently
described by $L$ in Eq. (\ref{eq:L in terms of omega}) to the
first order in $v$, then the force correction $ \delta \dot m_{ij}
\dot q_j $ and torque correction $\delta \dot I_{ij} \omega_j$
should be small compared with that generated by the $- \dot{\bf q}
\cdot {\bf \Lambda} - {\boldsymbol \omega}\cdot{\bf \Gamma}$ term.
Such a condition depends on the physical configuration of the
system. In the single mode situations with nonzero ${\bf
\Lambda}$ and ${\bf \Gamma}$ (see Sec. IV), one can observe that
$\delta \dot m_{ij}$ and $\delta \dot I_{ij}$ oscillate rapidly at
the field frequency, hence their effects can be averaged out and
become negligible in the spirit of the rotating wave approximation.

\section{Hamiltonian and quantization}

The Hamiltonian defined from $L$ of the system is given by
\begin{equation}
H \equiv {\bf p}\cdot \dot{\bf q} +
\sum_{\zeta=\alpha,\beta,\gamma} \pi_\zeta \dot{\zeta}  + \sum_k
P_k \dot{Q}_k - L, \label{eq: H_def}
\end{equation}
where $\bf p$, $P_k$, and $\pi_\zeta $ are canonical momenta
conjugate to ${\bf q}$, $Q_k$, and the Euler angle $\zeta$
respectively. However, since it is rather inconvenient to handle Euler
angles, we introduce a {\em canonical angular momentum} defined by
${\bf J} = (\partial L/\partial {\boldsymbol \omega})$, which is
related to $\pi_\zeta$ via the relation: ${\bf J}\cdot
{\boldsymbol \omega} = \dot{\alpha} \pi_\alpha + \dot{\beta}
\pi_\beta + \dot{\gamma} \pi_\gamma$. From the Lagrangian, $\bf
p$, $\bf J$,  and $P_k$ are given by
\begin{eqnarray}
{\bf p} &=& m \dot{\bf q}
- \sum_{kj} {\boldsymbol \eta}_{kj} \dot{Q}_k Q_j,
\label{eq: p def} \\
{\bf J} &=&  I {\boldsymbol \omega} - \sum_{kj} {\bf g}_{kj}
\dot{Q}_k Q_j,
\label{eq: J_def}\\
P_k &=& \dot{Q}_k - \sum_j \left(\dot{\bf q}\cdot {\boldsymbol
\eta}_{kj} + {\boldsymbol \omega}\cdot {\bf g}_{kj}\right) Q_j.
\label{eq: Pk_def}
\end{eqnarray}
Note that $\bf p$ and $\bf J$ differs from the kinetic momentum
$m\dot{\bf q}$ and angular momentum $I {\boldsymbol \omega}$ for
nonzero fields, respectively.

The solutions of $\dot {\bf q}$, ${\boldsymbol \omega}$ and $\dot
Q_k$ in terms of the canonical momenta are complicated because of
the coupled equations (\ref{eq: p def})--(\ref{eq: Pk_def}). However, since
field-dependent terms that are quadratic in ${\bf v}$ have been
discarded the Lagrangian, we have to maintain the consistency by
dropping terms of same order in eliminating $\dot {\bf q}$,
${\boldsymbol \omega}$ and $\dot Q_k$ for the Hamiltonian.
Specifically, we shall write
\begin{eqnarray} && m \dot{\bf q}=
{\bf p} + \sum_{kj} {\boldsymbol \eta}_{kj} \dot{Q}_k Q_j \approx
{\bf p}
+ \sum_{kj} {\boldsymbol \eta}_{kj} P_k Q_j,
\label{eq: p approx}\\
&& I {\boldsymbol \omega}= {\bf J} + \sum_{kj} {\bf g}_{kj}
\dot{Q}_k Q_j \approx {\bf J} + \sum_{kj} {\bf g}_{kj} P_k Q_j
\label{eq: Pk approx}.
\end{eqnarray}
By substituting Eqs.~(\ref{eq: p approx}) and (\ref{eq: Pk approx}) into Eq. (\ref{eq: H_def}) and again neglecting
field corrected inertia contributions, we obtain an explicit
expression of the Hamiltonian (\ref{eq: H_def}),
\begin{eqnarray}
H =  \frac{\left({\bf p} + {\bf \Lambda}'\right)^2}{2m} +
\frac{\left({\bf J} + {\bf \Gamma}'\right)^2}{2I} + \frac{1}{2}
\sum_{k} \left[P_k^2 + \omega_k^2({\bf q})Q_k^2\right],
\label{eq: H PQ}
\end{eqnarray}
where ${\bf \Lambda}' = \sum_{kj} {\boldsymbol \eta}_{kj} P_k Q_j$ and ${\bf \Gamma}'  = \sum_{kj} {\bf
g}_{kj} P_k Q_j$ are defined.

It is interesting that this Hamiltonian takes a form similar to
the minimal-coupling Hamiltonian in electrodynamics, with ${\bf
\Lambda}'$ and ${\bf \Gamma}'$ somehow playing the role of vector
potential in the kinetic energy term. In addition, the Hamiltonian
indicates that there is a rotational coupling of the sphere
described by the second term in (\ref{eq: H PQ}). It is useful
to rewrite ${\bf \Gamma}'$ as
\begin{eqnarray}
{\bf \Gamma}' = -  \int d^3r
\left(\frac{\epsilon-1}{\epsilon}\right)
            \left[({\bf r}-{\bf q})\times\left({\bf \Pi}\times{\bf B}\right)\right] ,
\end{eqnarray}
where ${\bf B} = \sum_k Q_k (\nabla\times{\bf u}_k)$ and ${\bf
\Pi} =  \epsilon \sum_k P_k{\bf u}_k$ are the magnetic field and
field canonical momentum density, respectively. The form of ${\bf
\Gamma}'$ is very similar to the field angular momentum stored in
the dielectric, apart from a proportionality constant. Therefore,
approximately speaking, the second term of Hamiltonian
(\ref{eq: H PQ}) represents an angular momentum coupling, i.e.,
the interaction corresponds to an exchange of angular momenta
between the field and the sphere. Furthermore, since ${\bf
\Gamma}'({\bf q})$ depends on ${\bf q}$, there is also a coupling between the mechanical rotation and c.m. motion of the sphere,
mediated by the fields.

With the classical Hamiltonian (\ref{eq: H PQ}), the canonical
quantization of the system is readily achieved by postulating the
dynamical variables into operators with the commutation relations:
$[q_{\mu},p_{\nu}] = i\hbar \delta_{\mu\nu}$, $[J_{\mu}, J_{\nu}]
= i\hbar \epsilon_{ \mu \nu \kappa} J_{\kappa} $, $[K_{\mu}, K_{\nu}]
= -i\hbar \epsilon_{ \mu \nu \kappa} K_{\kappa} $, and  $[Q_k, P_j]
= i\hbar \delta_{kj}$, where the Greek subscripts refer to the three
axes in rectangular coordinates, and $K_\mu$ are the body-axis
components of $\bf J$ \cite{euler ang}. In this way the quantum
Hamiltonian of the system takes the same expression as
Eq.~(\ref{eq: H PQ}), but with $P_k Q_j$ symmetrized by
$\left(P_k Q_j+Q_jP_k \right) /2$ in ${\bf \Lambda}'$ and ${\bf
\Gamma}'$.

In order to represent photon states of the system, we introduce
the ${\bf q}$-dependent annihilation and creation operators for each
cavity field mode:
\begin{eqnarray}
a_{k}({\bf q}) &=& \sqrt{\frac{1}{2\hbar \omega_{k}({\bf q})}}
\left[\omega_{k}({\bf q}) {Q}_{k} + i {P}_{k}\right]
\label{eq:a(q) def},\\
a^{\dag}_{k}({\bf q}) &=& \sqrt{\frac{1}{2\hbar \omega_{k} ({\bf
q})}}\left[\omega_{k}({\bf q}) {Q}_{k} - i {P}_{k}\right]
\label{eq:a^dag(q) def},
\end{eqnarray}
which satisfy the commutation relation $[a_{k}({\bf q}),
a_{j}^{\dag}({\bf q})] = \delta_{kj}$. Since $a_{k}({\bf q})$
depends on ${\bf q}$, for each position of the dielectric we have
a set of Fock states associated with that position. These states
can be labeled as $|\{n_{k}\}, {\bf q}, \xi \rangle$, where
$\{n_{k}\}=\{n_{1},n_{2},n_{3},\ldots\}$ denotes the occupation
number of each photon mode, and $\xi = (j,m,k)$ denotes the
eigenbasis vectors of $\bf J$ (see \cite{euler ang}):
\begin{eqnarray}
{\bf J}^2 |\{n_{k}\}, {\bf q}, \xi \rangle
&=& \hbar^2 j (j+1) |\{n_{k}\}, {\bf q}, \xi \rangle,\\
J_z |\{n_{k}\}, {\bf q}, \xi \rangle
&=& \hbar m |\{n_{k}\}, {\bf q}, \xi \rangle,\\
K_z |\{n_{k}\}, {\bf q}, \xi \rangle
&=& \hbar k |\{n_{k}\}, {\bf q}, \xi \rangle,
\end{eqnarray}
where $K_z$ is the $z$ component of $\bf J$ in the body coordinates.
Here $|\{n_{k}\}, {\bf q}, \xi \rangle$ is a simultaneous
eigenstate of the photon-number operator $a^{\dag}_{k}({\bf
q})a_{k}({\bf q})$ and the position operator ${\bf q}$ i.e.,
\begin{eqnarray}
&& a^{\dag}_{k}({\bf q})a_{k}({\bf q})|\{n_{k}\}, {\bf q},
\xi \rangle = n_{k} |\{n_{k}\}, {\bf q}, \xi \rangle, \\
&& \hat{\bf q} |\{n_{k}\},{\bf q}, \xi \rangle = {\bf q} |\{n_{k}\},
{\bf q}, \xi \rangle.
\end{eqnarray}
Such a set of eigenstates is orthonormal and complete, so that any
quantum state of the whole system $|\Psi\rangle$ can be expanded
in the basis of these eigenstates, i.e.,
\begin{equation}
|\Psi\rangle = \sum_{\xi, \{n_{k}\}} \int d^3{\bf q}
C(\{n_{k}\},{\bf q}, \xi ) |\{n_{k}\}, {\bf q}, \xi \rangle ,
\end{equation}
where $C(\{n_{k}\}, {\bf q}, \xi )$ is the probability amplitude.

With the help of the ${\bf q}$-dependent annihilation and creation
operators, the Hamiltonian Eq.~(\ref{eq: H PQ}) becomes
\begin{equation}
H =  \frac{\left({\bf p} + {\bf \Lambda}'\right)^2}{2m} +
\frac{\left({\bf J} + {\bf \Gamma}'\right)^2}{2I} +\sum_{k} \hbar
\omega_{k}({\bf q}) \left(a^{\dag}_{k} a_{k} +\frac{1}{2}\right),
\label{eq: H fock}
\end{equation}
where we have used a shorthand $a_{k} = a_{k}({\bf q})$ for
convenience, and
\begin{eqnarray}
{\bf \Lambda}'({\bf q}) &=& -\frac{i\hbar}{2} \sum_{k,j}
{\boldsymbol \eta}_{kj}({\bf q}) \sqrt\frac{\omega_k({\bf
q})}{\omega_j({\bf q})} \left( a_k a_j - a_k^{\dag} a_j^{\dag} +
a_k^{\dag} a_j - a_j^{\dag} a_k \right),
\label{eq: Lambda a a dag} \\
{\bf \Gamma}'({\bf q}) &=& -\frac{i\hbar}{2} \sum_{k,j} {\bf
g}_{kj}({\bf q}) \sqrt\frac{\omega_k({\bf q})}{\omega_j({\bf q})}
\left( a_k a_j - a_k^{\dag} a_j^{\dag} + a_k^{\dag} a_j -
a_j^{\dag} a_k \right) \label{eq: g a a dag}.
\end{eqnarray}
Note that ${\bf \Lambda}'$ and ${\bf \Gamma}'$ contains
photon-number nonconserving terms $a_k^{\dag} a_j^{\dag}$ which
are responsible for photon generation in the dynamical Casimir
effect \cite{Dodonov}, but this is a subject beyond the scope of
this paper. For fields at optical frequencies, the $a_k^{\dag}
a_j^{\dag}$ terms are fast oscillating in the interaction picture,
and so in the spirit of rotating wave approximation, only the photon-number
conserving terms $a_k^{\dag} a_j$ will be kept in ${\bf \Lambda}'$
and ${\bf \Gamma}'$. These terms describe the scattering of
photons between different modes due to the motion of the sphere.

\section{Single-mode situations}

In this section, we discuss the Hamiltonian under the single-mode
approximation. This applies to situations when the field is
dominantly contributed by a single mode $k$, and the scattering of
photons from the $k$ mode to other modes is negligible within a
coherent interaction time. From Eqs. (\ref{eq: Lambda a a dag}) and
(\ref{eq: g a a dag}), it follows that under the single-mode
consideration, ${\bf \Lambda}'$ and ${\bf \Gamma}'$ only contain
photon-number nonconserving terms $a^{\dag 2}_{k}$ and $a_{k}^2$,
and vanish under the rotating wave approximation. Hence the
corresponding Hamiltonian reads
\begin{equation}
H \approx \frac{{\bf p}^2}{2m} + \frac{{\bf J}^2}{2I} + \hbar
\omega_{k}({\bf q}) a^{\dag}_{k} a_{k}, \label{eq: adiabatic H}
\end{equation}
in which the rotational motion of the sphere is decoupled from the
field, and the optomechanical coupling appears only through the
position dependent mode frequency $\omega({\bf q})$. Such a form
of the Hamiltonian has been considered in Refs. \cite{kimble
nanosphere,cirac}.

However, we emphasize that the single-mode Hamiltonian (\ref{eq:
adiabatic H}) is based on the real mode function ${\bf u}_k$ in
the derivation. The situation is different if complex mode
functions are involved, for example, in a ring cavity which supports traveling wave modes. In the Appendix, we show how the Hamiltonian (\ref{eq: H fock}) can be modified to incorporate complex modes, in which a complex mode function is formed by a linear combination of real modes of the same frequency. In particular,
when photons mainly occupy a complex mode ${\bf f}$, by the single mode approximation the Hamiltonian becomes
\begin{equation}
H =  \frac{\left({\bf p}+ {\boldsymbol \lambda} b^{\dag}b
\right)^2}{2m} + \frac{\left({\bf J} + {\boldsymbol \gamma}
b^{\dag}b  \right)^2}{2I} + \hbar \omega ({\bf q}) \left(b^{\dag}
b+\frac{1}{2}\right), \label{eq: single}
\end{equation}
where $b$ and $\omega({\bf q})$ are the annihilation operator and mode frequency associated with the $\bf f$ mode, respectively, and ${\boldsymbol \lambda}({\bf q})$ and ${\boldsymbol \gamma}({\bf q})$ are coupling strengths determined by the mode function,
\begin{eqnarray}
 && {\boldsymbol \lambda}({\bf q}) = -\hbar \Im \int d^3r
\left[\epsilon
\sum_{l=x,y,z}
\left({\bf f}^* \cdot \hat{\bf e}_l\right)
\nabla_{\bf q} \left({\bf f} \cdot \hat{\bf e}_l\right)
+
\left(\epsilon-1\right) {\bf f}^* \times\left(\nabla\times{\bf
f}\right) \right],
\label{eq: single lambda}
\\
&& {\boldsymbol \gamma}({\bf q})= -\hbar \Im \int d^3r
\left(\epsilon-1\right) \left({\bf r}-{\bf
q}\right)\times\left[{\bf f}^* \times\left(\nabla\times{\bf
f}\right) \right].
\label{eq: single gamma}
\end{eqnarray}
We see that the velocity dependent coupling reappears
in the single complex mode.
Physically, ${\boldsymbol \lambda}({\bf q})$ and ${\boldsymbol \gamma}({\bf q})$ can roughly be understood as a measure of the field momentum and angular momentum stored in the sphere, contributed by a $\bf f$ mode photon. A complex mode, such as that of a traveling wave, can carry net field momentum (angular momentum).
On the other hand, a real mode can be decomposed as a linear combination of complex modes, whose contributions of momenta (angular momenta) cancel each other.
A familiar example is the standing wave mode function $\sin{( {\bf k} \cdot {\bf r})}$,
which is a superposition of two traveling wave modes $e^{i {\bf k} \cdot {\bf r}}$ and $e^{-i {\bf k} \cdot {\bf r}}$.
This explains why the velocity dependent coupling only appears in the complex mode situation.

To estimate the magnitude of ${\boldsymbol \lambda}({\bf q})$
and ${\boldsymbol \gamma}({\bf q})$, we consider a physical situation where a dielectric sphere of subwavelength size is placed in a ring cavity, and the field excitation is dominantly contributed by a Gaussian beam of an optical tweezer.
The configuration of ring cavity supports traveling wave modes,
so that ${\boldsymbol \lambda}({\bf q})$ and ${\boldsymbol \gamma}({\bf q})$ can be nonvanishing.
Under the paraxial approximation, the field mode function is given by,
\begin{eqnarray}
&&{\bf f}(x,y,z) = \left[u(x,y,z)\hat{\bf e}_x + \frac{i}{k} \frac{\partial u}{\partial x}\hat{\bf e}_z\right] \frac{e^{ikz}}{\sqrt L_c},
\label{eq: f}\\
&&u(x,y,z) = \sqrt{\frac{2}{\pi}} \frac{1}{w(z)}
\exp\left[-\frac{x^2+y^2}{w^2(z)} \left(1- i \frac{z}{z_R}\right)\right]
\exp\left[-i\tan^{-1}\left(\frac{z}{z_R}\right)\right],
\label{eq: u}
\end{eqnarray}
where $L_c$ is the effective length of the cavity.
The beam is linearly polarized in $\hat{\bf e}_x$,
propagating along its wavevector ${\bf k} = k\hat{\bf e}_z$,
with a beam radius $w(z) = \sqrt{2(z^2+z_R^2)/k z_R}$ (focal plane at $z=0$), and $z_R$ is the Rayleigh range.
For a subwavelength sphere satisfying $kR\ll 1$ and $R/z_R \ll 1$, the Gaussian beam should be a good approximation to the normal-mode,
and the $\bf q$ dependence of the mode function should be weak such that the contribution of the term involving $\nabla_{\bf q}$ in
Eq.~(\ref{eq: single lambda}) is negligible, i.e.,
\begin{eqnarray}
{\boldsymbol \lambda}({\bf q}) \approx -\hbar \Im \int d^3r
\left(\epsilon-1\right) {\bf f}^* \times\left(\nabla\times{\bf
f}\right).
\label{eq: approx lambda}
\end{eqnarray}
Therefore ${\boldsymbol \lambda}$ and ${\boldsymbol \gamma}$ can be determined
by substituting Eqs.~(\ref{eq: f}) and (\ref{eq: u}) into Eqs.~(\ref{eq: single gamma}) and (\ref{eq: approx lambda}).
In particular, near the beam focus where their magnitudes are largest, we find
\begin{eqnarray}
{\boldsymbol \lambda}
&=& -\frac{4}{3} \frac{\hbar(n^2-1)}{L_c}
\left(\frac{R}{z_R}\right)^3
(k z_R)^2 \hat{\bf e}_z, \\
{\boldsymbol \gamma}
&=& -\frac{4}{15} \frac{\hbar(n^2-1) z_R }{L_c}
\left(\frac{R}{z_R}\right)^5
(k z_R)^2 (1+2k z_R)
\left(-\frac{q_y}{z_R} \hat{\bf e}_x
+ \frac{q_x}{z_R}\hat{\bf e}_y\right).
\end{eqnarray}
Numerical calculations were performed based on the parameters in Ref.~\cite{cirac}, where $n=1.45$, $R = 100\mbox{ nm}$, $L_c = 4\mbox{ mm}$, $\lambda = 2\pi/k = 1064 \mbox{ nm}$ and $z_R = 0.53 \mbox{ $\mu$m}$ for the optical tweezer at a power $P = 15\mbox{ mW}$ (corresponds to an average photon number $\langle b^{\dag}b\rangle \approx 10^6$). We find that if the sphere moves under thermal fluctuations at room temperature, the optical phase shift $\hbar^{-1}\left(\dot{\bf q}\cdot {\boldsymbol \lambda} + {\boldsymbol \omega}\cdot {\boldsymbol \gamma}\right)\Delta t$ accumulated within the coherent time scale $\Delta t \sim 0.1 \mbox{ ms}$ would be on the order $5.1\times10^{-5} \mbox{ rad}$, which is very small. Furthermore, under the strong coherent field of the optical tweezer, we find that the magnitude of the nonadiabatic force ${\bf F} = \dot{\bf q}\times(\nabla\times{\boldsymbol \lambda}) - \nabla({\boldsymbol \omega}\cdot{\boldsymbol \gamma})$ is negligible compared with the restoring force of the optical tweezer.

\begin{figure}[h]
\centering
\includegraphics[width= 8 cm]{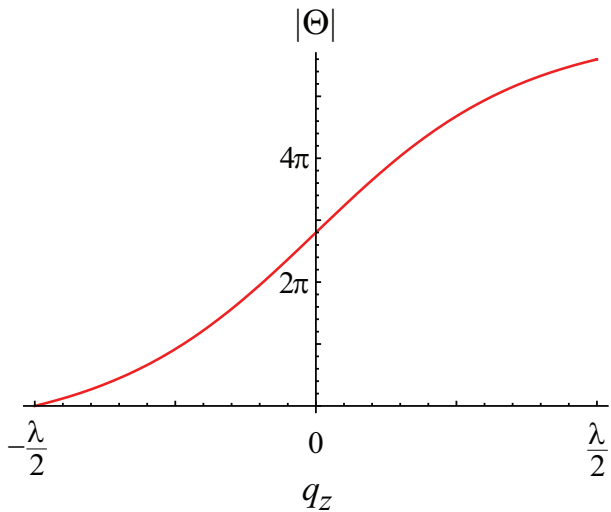}
\caption{Magnitude of geometrical phase $|\Theta|$ accumulated as the dielectric sphere travels from $q_z = -\lambda/2$ to $q_z = \lambda/2$ along the beam axis. The beam focus is at $q_z = 0$. We follow the numerical parameters as in Ref.~\cite{cirac}, where $n=1.45$, $R = 100\mbox{ nm}$, $L_c = 4\mbox{ mm}$, $\lambda = 1064 \mbox{ nm}$, and $z_R = 0.53 \mbox{ $\mu$m}$ for the optical tweezer at a power $P = 15\mbox{ mW}$ (hence $\langle b^{\dag}b\rangle \approx 10^6$).}
\label{fig2}
\end{figure}

While the velocity dependent effects on the subwavelength sphere appears to be quite weak under the single mode field,
we should point out that even under adiabatic motion of the sphere,
there can be an appreciable effect of geometrical phase due to ${\boldsymbol \lambda}$.
Let us suppose that the wave packet of the sphere travels from ${\bf q}_i$ to ${\bf q}_f$ along a path $C$,
under a strong coherent field (e.g., of the optical tweezer).
In this case, we may take the field state as classical by replacing $b^{\dag}b$  by its expectation value $\langle b^{\dag}b\rangle$. The minimal coupling due to ${\boldsymbol \lambda}$ then affects
the wave function of the sphere $\psi({\bf q})$ by attaching to it a path-dependent,
quantum mechanical (geometrical) phase,
apart from an overall dynamical phase:
\begin{eqnarray}
\psi({\bf q}_i) \rightarrow e^{-i\Theta}\psi({\bf q}_f),
\hspace{5mm}
\Theta = \hbar^{-1} \int_{C} d{\bf q}\cdot
{\boldsymbol \lambda}({\bf q}) \langle b^{\dag}b\rangle.
\end{eqnarray}
In particular, if the sphere travels along the beam axis of the optical tweezer,
\begin{eqnarray}
\Theta
%= -\frac{4}{3}(n^2-1)\frac{k^2 R^3}{L_c} \langle b^{\dag}b\rangle\int^{q_{z,f}}_{q_{z,i}} \frac{d (q_z/z_R)}{\left[1+(q_z/z_R)^2\right]^2} \nonumber \\
= -\frac{2}{3}(n^2-1)
\frac{k^2 R^3}{L_c} \langle b^{\dag}b\rangle
\left[\frac{(q_z/z_R)}{1+(q_z/z_R)^2} + \tan^{-1}\left(\frac{q_z}{z_R}\right)\right]^{q_{z_f}}_{q_{z_i}}.
\end{eqnarray}
Hence under the numerical parameters used above,
the sphere accumulates a phase of $\Theta \approx -5.4 \pi$ by traveling {\em one optical wavelength} across the beam focus (i.e. from $z=-\lambda/2$ to $z=\lambda/2$) (see Fig.~\ref{fig2}).
Note that since $\boldsymbol \lambda$ is proportional to $|{\bf f}|^2$, the Gouy phase factor (i.e., $\exp[-i\tan^{-1}(z/z_R)]$) and the optical phase factor $e^{ikz}$ do not contribute to the calculation of $\Theta$.

We point out that in setups which make use of real modes \cite{cirac2}, we have $\Theta =0$,
but for other configurations involving complex modes,
the geometrical phase could be nonzero.
We also remark that for a subwavelength sphere, $\Theta$ is proportional to the time averaged Poynting vector integrated along the path $C$,
which is analogous to the case of an induced dipole \cite{dipole phase}. However, if the sphere size is appreciable relative to the field wavelength,
the change of mode structure as the sphere moves should be properly taken into account.
In our approach, this can be readily achieved by including the contribution of the first term involving $\nabla_{\bf q}$ in Eq.~(\ref{eq: single lambda}).

\section{Conclusion}
To conclude, we have presented a nonrelativistic Lagrangian-Hamiltonian formalism of a moving dielectric sphere interacting with radiation fields. We see that in this three-dimensional system, the sphere's c.m. degree of freedom $\bf q$ is no longer independent with the vector potential ${\bf A}$ under the generalized radiation gauge, and this poses an interesting subtlety in the theory. We have resolved this issue by making use of the instantaneous normal-mode projection to consistently identify the independent degrees of freedom subject to the gauge, enabling canonical quantization of the system in the usual manner.

By including the sphere-field interaction up to first order in $v$, our Hamiltonian (\ref{eq: H PQ}) and (\ref{eq: H fock}) should capture velocity-dependent optomechanical processes that are not described under the adiabatic approximation. For example, coupling between the field (normal) modes can result from both the translational and rotational motion of the sphere through the interaction described by ${\boldsymbol \Lambda}'$ and ${\boldsymbol \Gamma}'$, respectively. In addition, these two mechanical degrees of freedom become coupled in the presence of radiation fields, due to the coupling characterized by ${\boldsymbol \Gamma}'({\bf q})$. Such motion-induced coupling can become significant in nonadiabatic regimes, especially when the oscillation (or rotation) frequency of the dielectric sphere is close to the frequency spacing between two field modes, in which case transitions between the two modes can be resonantly enhanced.

Under the single mode approximation, we have shown that the velocity dependent effects are typically very weak for a subwavelength sphere, hence it may justify some of the approximations in Refs.~\cite{kimble nanosphere, cirac}.
Nonetheless, even under such limiting considerations, we have also indicated an appreciable geometrical phase (in excess of $\pi$)
acquired by the sphere wavepacket as it moves adiabatically
under the single (complex) mode field.
Moreover, we should emphasize that our theory does not require the single-mode adiabatic approximation. With the explicit form of interaction strengths between the sphere and various field modes
in Eq.~(\ref{eq: H fock}), our work here enables a further study on the quantum dynamics and applications in multimode nonadiabatic regimes.

\emph{Acknowledgment} -- This work was partially supported by a
grant from the Research Grants Council of Hong Kong, Special
Administrative Region of China (Project No.~CUHK401810).

\appendix
\section {The complex mode Hamiltonian}
\label{complex}
The complex mode annihilation operators $\{b_k({\bf q}) \}$ are constructed as a linear combination of the (degenerate) real mode operators $\{a_k({\bf q})\}$ by
\begin{equation}
b_k({\bf q}) = \sum_j U_{kj} a_j({\bf q}),
\label{eq: b def}
\end{equation}
where the (complex) coefficients $U_{kj}$ satisfy the unitary property $\sum_j U_{kj}U^{*}_{k'j} = \sum_j U_{jk}U^{*}_{jk'} = \delta_{kk'}$, so that the commutation relation $[b_k({\bf q}), b_{k'}^{\dag}({\bf q})] = \delta_{kk'}$ is readily fulfilled.
Furthermore, since $b_k({\bf q})$ only mixes degenerate real modes, we have $U_{kj} = 0$ except $\omega_k({\bf q}) = \omega_j({\bf q})$.
The complex mode function associated with $b_k({\bf q})$ is given by
\begin{equation}
{\bf f}_k({\bf r},{\bf q})
= \sum_{j} U_{kj}^{*} {\bf u}_j({\bf r},{\bf q}),
\end{equation}
so that in terms of the vector potential ${\bf A}({\bf r})$
and the field canonical momentum density ${\bf \Pi}({\bf r})$,
\begin{equation}
b_k({\bf q}) = \sqrt{\frac{1}{2\hbar\omega_k({\bf q})}}
\int d^3 r
\left[
\epsilon({\bf r},{\bf q})\omega_k({\bf q}){\bf A}({\bf r})
+ i{\bf \Pi}({\bf r})
\right]
\cdot {\bf f}^{*}_k({\bf r},{\bf q}).
\end{equation}
Using Eq.~(\ref{eq: b def}) and the properties of $U_{kj}$,
we can rewrite the Hamiltonian (\ref{eq: H fock}) as
\begin{equation}
H =  \frac{\left({\bf p} + {\bf \Lambda}'\right)^2}{2m} +
\frac{\left({\bf J} + {\bf \Gamma}'\right)^2}{2I} +\sum_{k} \hbar
\omega_{k}({\bf q}) \left(b^{\dag}_{k} b_{k} +\frac{1}{2}\right),
\end{equation}
where ${\bf \Lambda}'$ and ${\bf \Gamma}'$ reads
\begin{eqnarray}
{\bf \Lambda}'({\bf q}) &=& -\frac{i\hbar}{2} \sum_{k,j}
 \sqrt\frac{\omega_k({\bf
q})}{\omega_j({\bf q})}
\left[{\boldsymbol \eta}^{(1)}_{kj}({\bf q}) b_k b_j
+ {\boldsymbol \eta}^{(2)}_{kj}({\bf q}) b_k^{\dag} b_j
-\mbox{H.c.}\right],
\label{eq: complex Lambda}
\\
{\bf \Gamma}'({\bf q}) &=& -\frac{i\hbar}{2} \sum_{k,j}
 \sqrt\frac{\omega_k({\bf
q})}{\omega_j({\bf q})}
\left[{\bf g}^{(1)}_{kj}({\bf q}) b_k b_j
+ {\bf g}^{(2)}_{kj}({\bf q}) b_k^{\dag} b_j
-\mbox{H.c.}\right],
\label{eq: complex Gamma}
\end{eqnarray}
and the coefficients are given by
\begin{eqnarray}
{\boldsymbol \eta}^{(1)}_{kj}({\bf q}) &=& -\int d^3 r [ \epsilon
\sum_{l=x,y,z} \left({\bf f}_k \cdot \hat{\bf e}_l\right)
\nabla_{\bf q} \left({\bf f}_j \cdot \hat{\bf e}_l\right) +
\left(\epsilon-1\right) {\bf f}_k \times\left(\nabla\times{\bf
f}_j\right) ],
\\
{\boldsymbol \eta}^{(2)}_{kj}({\bf q}) &=& -\int d^3 r [ \epsilon
\sum_{l=x,y,z} \left({\bf f}^{*}_k \cdot \hat{\bf e}_l\right)
\nabla_{\bf q} \left({\bf f}_j \cdot \hat{\bf e}_l\right) +
\left(\epsilon-1\right) {\bf f}^{*}_k \times\left(\nabla\times{\bf
f}_j\right) ],
\\
{\bf g}^{(1)}_{kj}({\bf q}) &=& -\int d^3 r \left(\epsilon-1\right)
\left({\bf r}-{\bf q}\right)\times\left[{\bf f}_k
\times\left(\nabla\times{\bf f}_j\right) \right],
\\
{\bf g}^{(2)}_{kj}({\bf q}) &=& -\int d^3 r \left(\epsilon-1\right)
\left({\bf r}-{\bf q}\right)\times\left[{\bf f}^{*}_k
\times\left(\nabla\times{\bf f}_j\right) \right].
\end{eqnarray}
It follows that when the field excitation is dominantly contributed by a single complex mode ${\bf f}_k$, Eqs.~(\ref{eq: complex Lambda}) and (\ref{eq: complex Gamma}) can be reduced to, in the spirit of rotating wave approximation (where fast oscillating terms such as $b_k^2$ and $b^{\dag2}_k$ are neglected),
\begin{eqnarray}
{\bf \Lambda}'({\bf q})
&=& -\frac{i\hbar}{2}
\left[{\boldsymbol \eta}^{(2)}_{kk}({\bf q})
- {\boldsymbol \eta}^{(2)*}_{kk}({\bf q})\right]
b_k^{\dag} b_k
\equiv {\boldsymbol \lambda}({\bf q}) b^{\dag} b, \\
{\bf \Gamma}'({\bf q})
&=& -\frac{i\hbar}{2}
\left[{\bf g}^{(2)}_{kk}({\bf q})
- {\bf g}^{(2)*}_{kk}({\bf q})\right] b_k^{\dag} b_k
\equiv {\boldsymbol \gamma}({\bf q}) b^{\dag} b,
\end{eqnarray}
where the explicit form of ${\boldsymbol \lambda}({\bf q})$ and
${\boldsymbol \gamma}({\bf q})$ are given in
Eqs.~(\ref{eq: single lambda}) and (\ref{eq: single gamma}),
and $b=b_k({\bf q})$ and ${\bf f}= {\bf f}_k$ are used for shorter
notations. Hence the Hamiltonian (\ref{eq: single}) readily follows
in the limit of single complex mode.
As a remark, a single real mode can (effectively)
be decomposed into sums of degenerate complex modes whose contributions to the photon-number-conserving terms in Eqs.~(\ref{eq: complex Lambda}) and (\ref{eq: complex Gamma}) cancel out.

\end{document}